# Magnetic behavior of nanocrystalline $LaMn_2Ge_2$

S. Narayana Jammalamadaka, Sitikantha D. Das, B. A. Chalke and
E.V. Sampathkumaran[*]

*Tata Institute of Fundamental Research, Homi Bhabha Road, Colaba, Mumbai – 400005, India.*

The compound, $LaMn_2Ge_2$, crystallizing in $ThCr_2Si_2$-type tetragonal crystal structure, has been known to undergo ferromagnetic order below ($T_C=$) 326 K. In this article, we report the magnetic behavior of nanocrystalline form of this compound, obtained by high-energy ball-milling. $T_C$ of this compound is reduced marginally for the nanoform, whereas there is a significant reduction of the magnitude of the saturation magnetic moment with increasing milling time. The coercive field however increases with decreasing particle size. Thus, this work provides a route to tune these parameters by the reducing the particle size in this ternary family.

The rare-earth (R) compounds, crystallizing in ThCr$_2$Si$_2$-type layered tetragonal crystal structure have been attracting a lot of attention of condensed matter physicists for the past few decades due to various exotic phenomena exhibited by this family [1]. Among these, the Mn containing ones are of special interest, as other transition-metals, generally speaking, do not carry their own magnetic moment. The Mn-Mn interatomic distance along the basal plane falls in the range 2.8 - 3.2 Å, whereas the corresponding value along $c$-axis is nearly twice (5.4 – 5.6 Å). The nature of the magnetic structure has been known to sensitive to the Mn-Mn separation along the basal plane. For instance, in the RMn$_2$Ge$_2$ family, for most of the light R, ferromagnetic interaction is observed, whereas for heavy R, inter-layer antiferromagnetism (with intra-layer ferromagnetism) has been observed. Therefore, it is of interest to see how a reduction in particle size to nanoform influences the magnetic behavior in this family.

For our purpose, we have chosen LaMn$_2$Ge$_2$, in which there is no interference from the magnetism of rare-earth ions. This compound has been known to undergo ferromagnetic ordering around ($T_C=$ ) 326 K [1], which was also theoretically established [2]. It appears that, in the ferromagnetically ordered state, the Mn moments have a conical arrangement with a ferromagnetic component along the $c$-axis [3, 4]. The origin of canted ferromagnetism was addressed by electronic structure calculations and the value of the saturation magnetic moment ($\mu_s$) on Mn (3 $\mu_B$ per formula unit) could also be derived [5]. The most fascinating observation was that this compound is characterized by a large magnetoresistance at low temperatures, however with a *positive* sign, that too without tracking magnetization, uncharacteristic of ferromagnets [6]. This "inverse giant magnetoresistance effect", gave rise to some theoretical interest [6]. The magnetoresistance behavior of this natural multilayer is somewhat similar to that of artificial multilayer [7]. Thus, this compound is a sufficiently interesting magnetic system for investigations in the nanocrystalline form.

The compound in the ingot form was prepared by arc melting stoichiometric amounts of high purity (better than 99.9%) constituent elements in an arc furnace in an atmosphere of argon. After first melting, excess Mn was added to compensate for the loss while melting. After several melting, the total loss of the material do not exceed 1%. The sample thus synthesized was found to be single phase by x-ray diffraction (Cu K$_\alpha$) within the detection limit of x-ray diffraction (about 1%). In addition, the back-scattered scanning electron microscopic (SEM) pictures (obtained with JEOL JSM 840A) confirmed this conclusion and composition homogeneity was further ensured by energy dispersive x-ray (EDX) analysis. The ingot (called specimen **A**) was then subjected to a coarse-grinding in an agate pestle and mortar in an inert atmosphere. The grounded powder was then milled in a planetary ball mill (Fritsch pulverisette-7 premium line) with an operating speed of 800 rpm in a medium of toluene. Tungsten carbide vials and balls of 5 mm diameter were used with a balls-to-material mass ratio of 20:1. The specimens employed for investigations were the ones milled for 15 minutes (labeled ***B***), 45 minutes (labeled ***C***) and 3 hours (labeled **D**). XRD and SEM/EDX analysis at frequent intervals of time after completion of milling revealed that these powders are stable in air without oxidation for several days. Transmission electron microscope (Tecnai 200kV) and XRD were employed to determine the particle size. Magnetization (M) measurements (4.2 – 330 K) were carried out with the help of a commercial superconducting quantum interference device (Quantum Design, USA).

In figure 1, we show the x-ray diffraction pattern in a narrow angle range in the vicinity of most intense line. It is transparent from the figure that the lattice constants ($a$ = 4.195 Å, $c$ = 10.980 Å) do not change with increasing milling time within ± 0.010 Å, but the lines tend to broaden as the milling time increases. If one estimates the average particle size from Debye-Scherer relation, it is interesting to note that a milling for 15 mins is enough to reduce the particle size to about 40 nm. Further increase in milling time to 45 minutes reduces average particle size marginally, but after 3 hrs of milling, one obtains particles of about 30 nm. In fact, we have also taken TEM pictures on the particles obtained by ultrasonification of the specimens *C* and *D* to get a more precise idea of particle-size and the results (see figure 2) confirm the formation of nanoparticles with the particle size falling in the range 20 to 30 nm for *C* with the appearance of additional smaller particles for *D*.

We show the temperature dependent magnetization behavior in the range 5 – 330 K in figure 3a, measured in the presence of a field (H) of 5 kOe. There is a clear drop around 300 K for all specimens. In order to get an idea of magnetic transition temperature with reducing particle size, it is more informative to look at the plot of d*M*/dT, shown in figure 3b. There is a negative minimum at $T_C$ for the ingot, and therefore the temperature at which this occurs for various specimens is a measure of $T_C$. The value of $T_C$ thus inferred gets reduced gradually with increasing milling time, attaining a value close to 322, 318 and 314 K for specimens *B*, *C* and *D* respectively, though the transition appears to get broadened. At this moment, it may be recalled that the application of external pressure also reduces $T_C$ at the rate of -0.2 K/kbar [8]. Therefore, it is possible that the reduction in particle size acts like external pressure and the corresponding pressures deduced are about 20, 40, and 60 kbar respectively.

We now look at the isothermal magnetization, shown in figure 4, at different temperatures. The data at 300 K is also included for the sake of completeness, though it is close to $T_C$. As known earlier in the literature [2], for the parent specimen **A**, M exhibits a sudden increase for initial applications of H followed by a tendency for saturation at higher fields beyond 20 kOe. The value of $M_s$ for **A** at 5 K is about 3.35 $\mu_B$/formula-unit and decreases gradually with increasing temperature. For the finer particles, interestingly, there is a noticeable reduction as shown in figure 4 and in table 1. For instance, the value for **D** is as low as about 1.3 and 2.15 $\mu_B$/formula-unit at 200 and 5 K respectively. This implies that either there is a significant canting of ferromagnetic structure and/or there is a change in the itinerancy of Mn. There is also a gradual increase in the spontaneous magnetization from nearly 0.2 to 0.6 $\mu_B$/formula-unit (at 5 K) as one moves from **A** to **D**.

It is of interest to look at hysteresis behavior as well in isothermal M data (see figure 5). The curves are hysteretic even at room temperature. Since the $T_C$ is close to room temperature, it is not meaningful to compare the features at this temperature. It is transparent from this figure that, at any other lower temperature, the field-range over which hysteresis-loop is observed increases in the nanoparticles. The values of the coercive fields ($H_c$) are given in table 1. Thus, for instance, the value of $H_c$ increases from about 1.065 kOe to 5.274 kOe from particle **A** and **D**. Thus, a reduction in particle size enhances magnetic hardening [9].

Summarizing, there is a marginal reduction in the Curie temperature of the compound, $LaMn_2Ge_2$, in its nanocrystalline form. There is a dramatic change in the magnetic moment on Mn with decreasing particle size and thus this work provides a route to gradually vary the magnetic moment in this family. Though the saturation magnetic moment decreases with milling time, there is a significant enhancement in the coercive field with respect to the bulk form and therefore one can also tune the coercive field to a desirable value for any possible applications. We also hope this work triggers some theoretical efforts to work out a relationship between itinerant magnetism and particle size in this family.

**References**

*Corresponding author: e-mail address: sampath@mailhost.tifr.res.in
[1]   See, for instance, A. Szytula and J. Leciejewicz in "Handbook on the Physics and Chemistry of Rare Earths", edited by K.A. Gschneidner Jr. and L. Eyring (North-Holland, Amsterdam), **12,** p133 (1989), and references therein.
[2]   S. Ishida, S. Asano, and J. Ishida, J. Phys. Soc. Japan **55**, 936 (1986).
[3]   I. Nowik, Y. Levi, I. Felner, and E.R. Bauminger, J. Magn. Magn. Mater. **140-144,** 913 (1995); **147,** 373 (1995).
[4]   G. Venturini, B. Malaman, and E. Ressouche, J. Alloys Compd. 241, 135 (1996); G. Venturini, R. Welter, E. Ressouche, and B. Malaman, J. Magn. Magn. Mater. 150, 197 (1995). From these articles, it appears that the antiferromagnetic in-plane component persists well above Curie temperature till about 420 K.
[5]   S. Di Napoli, A.M. Llois, G. Bihlmayer and S. Blügel, Phys. Rev. B **75,** 104406 (2007); S. Di Napoli, S. Bihlmayer, S. Blügel, M. Alouani, H. Dreysee, and A.M. Llois, J. Magn. Magn. Mater. **272**, E265 (2004) and Physica B **354,** 154 (2004).
[6]   R. Mallik, E.V. Sampathkumaran, and P.L. Paulose, App. Phys. Lett. **71,** 2385 (1997); S. Majumdar, R. Mallik, E.V. Sampathkumaran, and P.L. Paulose, Solid State Commun. **108**, 349 (1998).
[7]   S. Di Napoli, S. Bihlmayer, S. Blügel, M. Alouani, H. Dreysee, and A.M. Llois, Physica B **354,** 154 (2004).
[8]   T. Kaneko, H. Yasui, T. Kanomata, and T. Suzuki, J. Magn. Magn. Mater. **104**, 1951 (1992).
[9]   During the course of this work, we came to know of a similar finding in the ball-milled $LaMn_2Si_2$: See, A. Elmali, S. Tekerek, I. Dincer, Y. Elerman, R. Theissmann, H. Ehrenberg, and H. Fuess, J. Magn. Magn. Mater. **320**, 364 (2008).

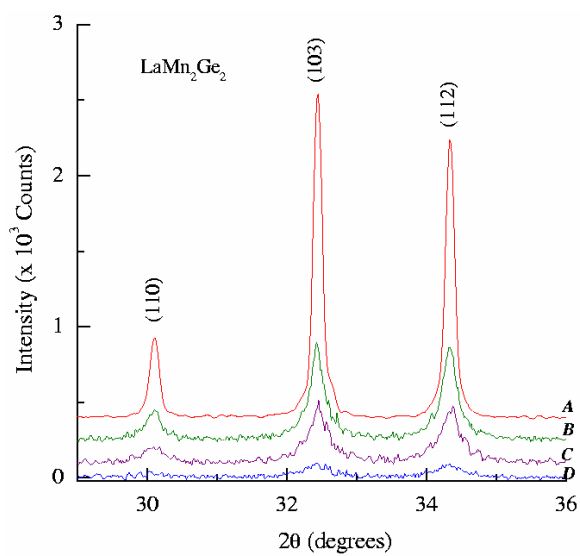

Figure 1:
(color online) X-ray diffraction patterns in a small angle range for LaMn$_2$Ge$_2$ for the parent specimen (*A*) and for the specimens obtained by milling for 15 min (*B*), 45 min (*C*) and 3 hrs *(D)*. The curves for *A, B* and *C* (plotted after subtracting the background) are shifted along y-axis for the sake of clarity.

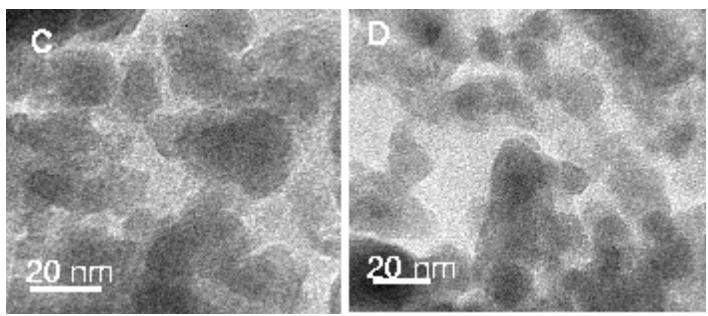

Figure 2:
TEM pictures of particles of specimens *C* and *D*.

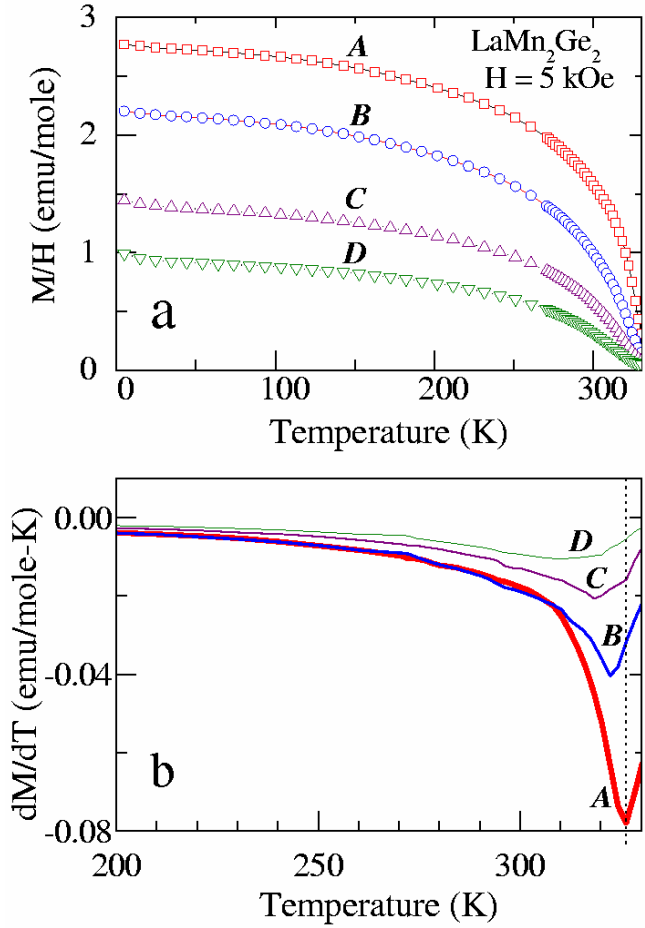

Figure 3:
(color online) (a) Magnetization divided by magnetic field obtained in a field of 5 kOe in the temperature range 4.2 to 300 K for the parent specimen, **A**, and for nanocrystals, **B**, **C**, and **D** of $LaMn_2Ge_2$. (b) The plots of dM/dT as a function of temperature.

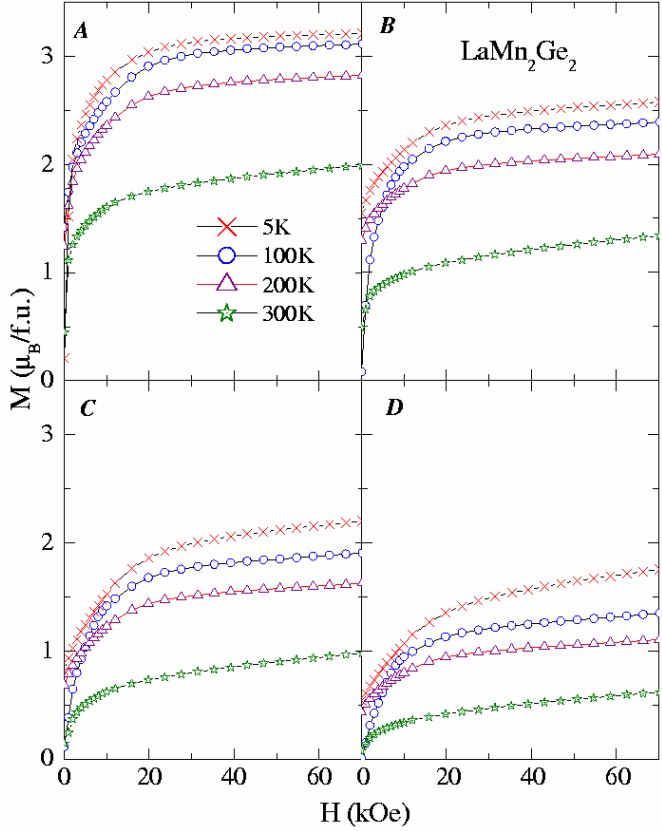

Figure 4:
Isothermal magnetization behavior of the parent specimen, **A**, and nanocrystals, **B**, **C**, and **D** of LaMn$_2$Ge$_2$ at various temperatures. The lines through the data points are guides to the eyes.

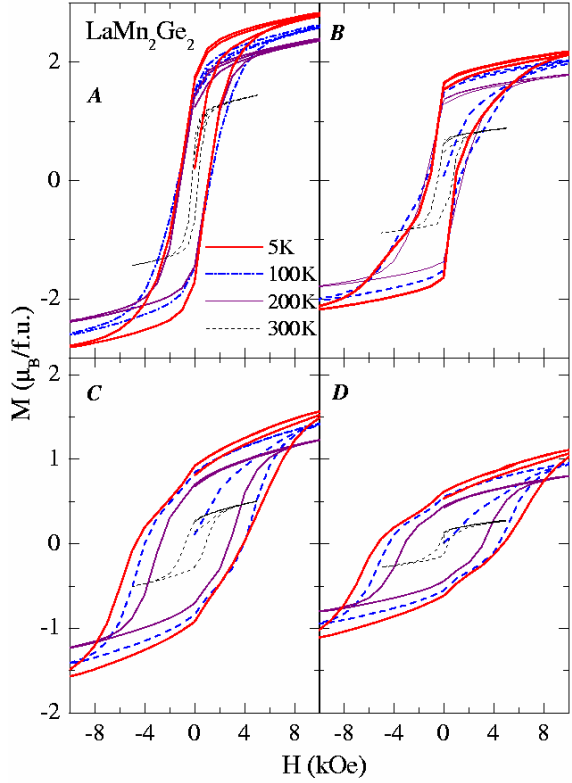

Figure 5:
Magnetic hysteresis behavior of the parent specimen, **A**, and nanocrystals, **B**, **C**, and **D** of LaMn$_2$Ge$_2$ at various temperatures.

**Table. 1**
The saturation moment and coercive fields (H$_C$) at various temperatures and Curie temperatures for specimens, **A, B**, **C** and **D,** of LaMn$_2$Ge$_2$.

| Specimen | M$_s$ ( ±0.05 µ$_B$/formula unit) | | | | H$_C$ (±5 Oe) | | | | T$_C$ (K) |
|---|---|---|---|---|---|---|---|---|---|
| | 5K | 100K | 200K | 300K | 5K | 100K | 200K | 300 K | |
| **A** | 3.35 | 3.25 | 2.95 | 2.25 | 1065 | 1065 | 937 | 245 | ~ 326 |
| **B** | 2.75 | 2.50 | 2.25 | 1.60 | 1515 | 1433 | 1199 | 447 | ~ 322 |
| **C** | 2.45 | 2.05 | 1.80 | 1.25 | 4766 | 3993 | 2764 | 684 | ~ 318 |
| **D** | 2.15 | 1.60 | 1.30 | 0.90 | 5274 | 4391 | 3141 | 288 | ~ 314 |